\documentclass[journal]{IEEEtran}
\usepackage{graphicx}
\usepackage{subfigure}
\usepackage{amsmath}

\begin{document}

\title{
  Perturbation Theory Based on Darboux Transformation on One-Dimensional Dirac Equation in Quantum Computation
}

\author{
  Agung Trisetyarso, \textit{Telkom Institute of Technology} \\
}

\maketitle

\begin{abstract}

We present the recent works \cite{trisetyarso2011} on the application of Darboux transformation on one-dimensional Dirac equation related to the field of Quantum Information and Computation (QIC). The representation of physical system in one-dimensional equation and its transformation due to the Bagrov, Baldiotti, Gitman, and Shamshutdinova (BBGS)-Darboux transformation showing the possibility admitting the concept of relativity and the trade-off of concurrent condition of quantum and classical physics play into the area of QIC. The applications in cavity quantum electrodynamics and on the proposal of quantum transistor are presented.

\end{abstract}


\IEEEpeerreviewmaketitle


\section{Results}
\IEEEPARstart{S}{everal} new proposals of perturbation theory in quantum mechanics are coined in order to answer the questions emerging in the recent research of quantum science and technology (in which, especially, related to quantum information and computation), such as proposed in quantum many body problems \cite{PhysRevLett.101.070503} and group theory \cite{trisetyarso2008degeneracy}. In this work, we present a new model of perturbation theory under the scheme of Darboux transformation \cite{bagrov1997darboux}\cite{trisetyarso2009application} acting upon one-dimensional Dirac equation:  intertwining operation on the \textit{old} Dirac Hamiltonian creates a new perturbing potential term on the  \textit{new} Dirac Hamiltonian. The advantage of this approach are: \textit{first}, the potential in Dirac Hamiltonian is in \textit{vector} form, instead of scalar form as conventionally used in Schr\"{o}dinger Hamiltonian. \textit{Second}, the Dirac potential can be a vector in three-dimensional Euclidean Bloch sphere or a \textit{complex four-vector} in Minkowskian Bloch sphere. \textit{Third}, under this scheme, the correlation between potential and physical quantities (atomic population, quantum state etc.) can be easily obtained. In other words, the physical quantities of a system is controllable via Dirac potential. Two folds of work are presented: \textit{first}, in the case of cavity quantum electrodynamics \cite{trisetyarso01} \cite{trisetyarso2011erratum}, and \textit{second}, in the case of a relativistic charge qubit in \textit{classical} external electromagnetic fields \cite{trisetyarso02}.  Below, we point out the outlines of our works.

\begin{figure}[p!]
\centering
\subfigure[]{
\includegraphics[scale=0.45]{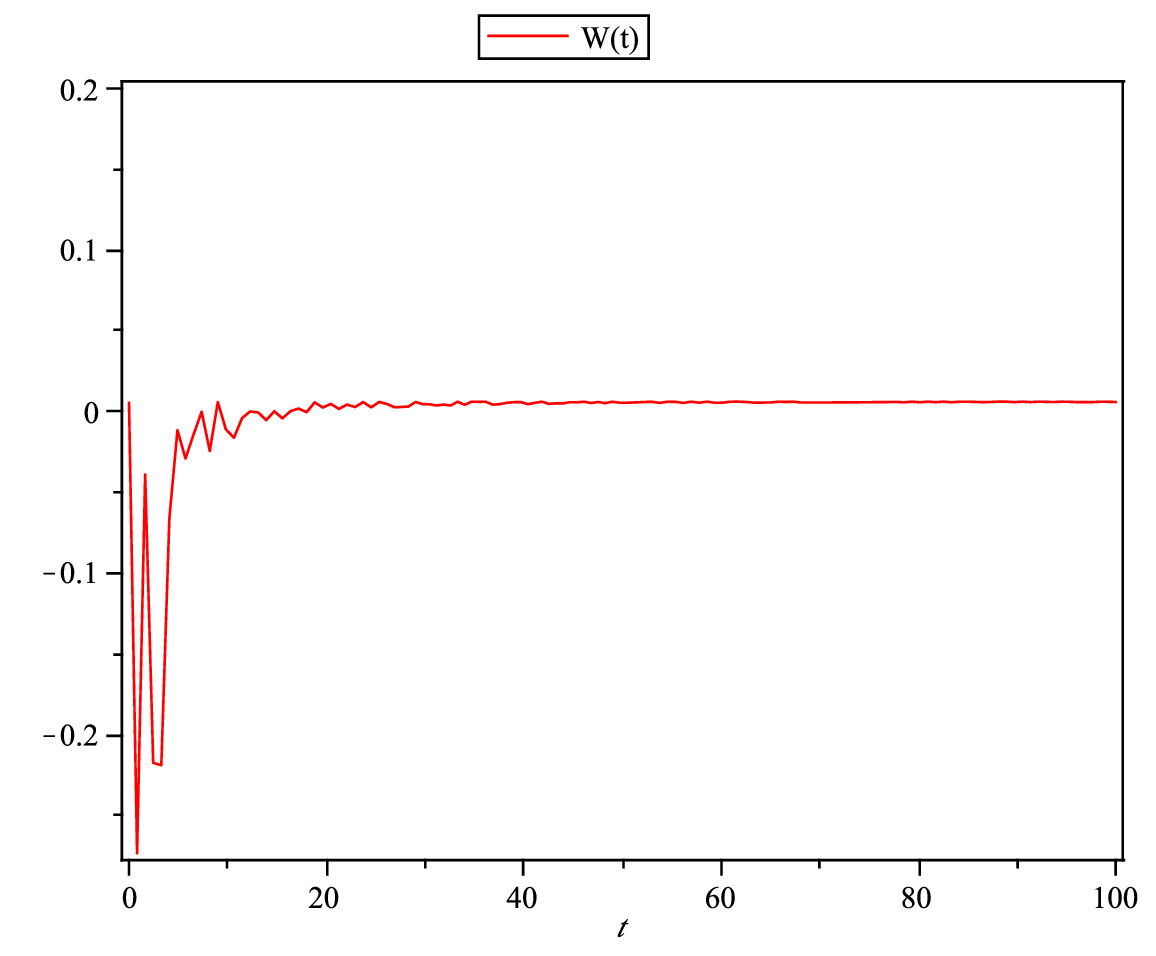}
\label{fig1}
}
\subfigure[]{
\includegraphics[scale=0.45]{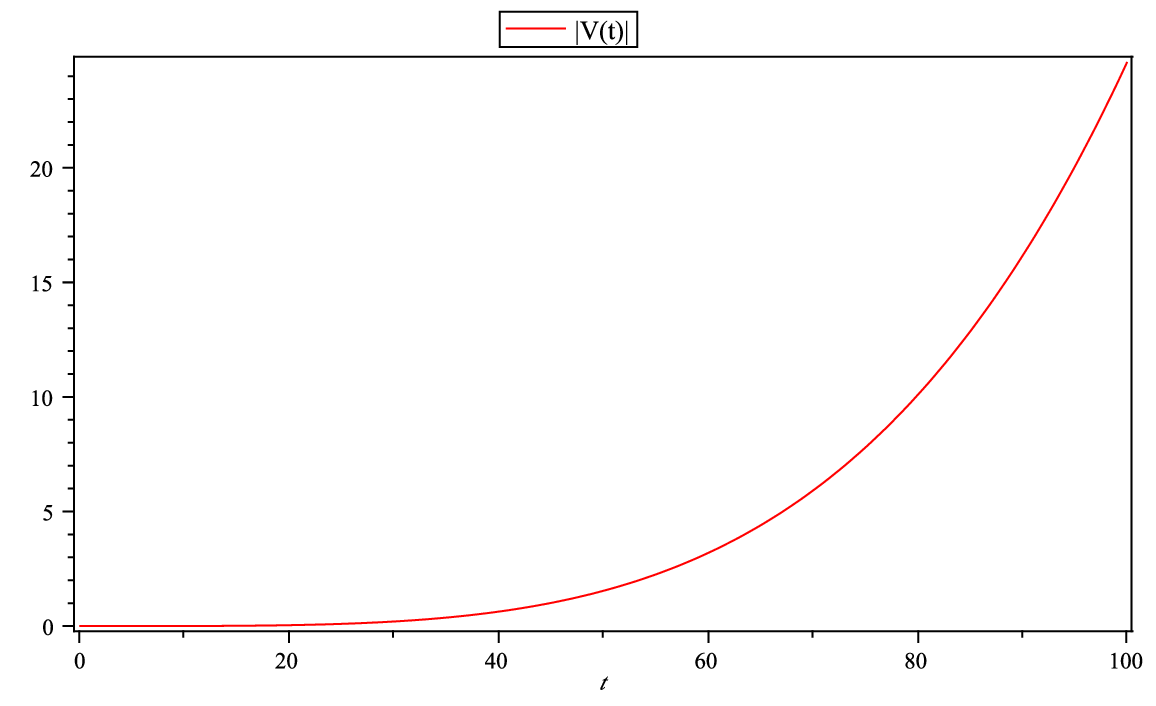}
\label{fig2}
}
\caption[f1]{\label{f1} The time evolution of the population of two states (\ref{fig1}) and the time evolution of Dirac potential (\ref{fig2}) in cavity quantum electrodynamics of ${\mathcal D(\sigma_{1})}$
for $t=100$.}
\end{figure}

The perturbation on a system of atom-photon in cavity quantum electrodynamics  due to the presence of the field behaving concurrently classical and quantum can be well-described by this scheme in which the potential is in a vector in three-dimensional Euclidean Bloch sphere. Under this scheme, the atomic population is controllable by choosing the Pauli matrices $\{\sigma_{i}|i=1,2,3\}$ as the input of BBGS-Darboux transformation, ${\mathcal D}(\sigma_{i})$. This operator acts as an \textit{controller} on an open-loop mechanism and the input-output of the system is a set of potential and state, $\{V,\Psi\}$, as stated in the Theorem 1 in the next Section. The benefit of  using this method is that the atomic inversion in cavity quantum electrodynamics is controllable via the Dirac potential. Especially, we found that the choice of ${\mathcal D}(\sigma_{1})$ causes the total collapse of Rabi oscillations. The remarkable feature of this choice is the appearance of wells prior to the total collapse which is a phenomenon similar to Yukawa potential thus it is so-called \textit{Yukawa-Rabi oscillations} which is produced by parabolic Dirac potential as shown in Fig. (\ref{f1}).

\begin{figure}[p!]
\includegraphics[scale=0.34]{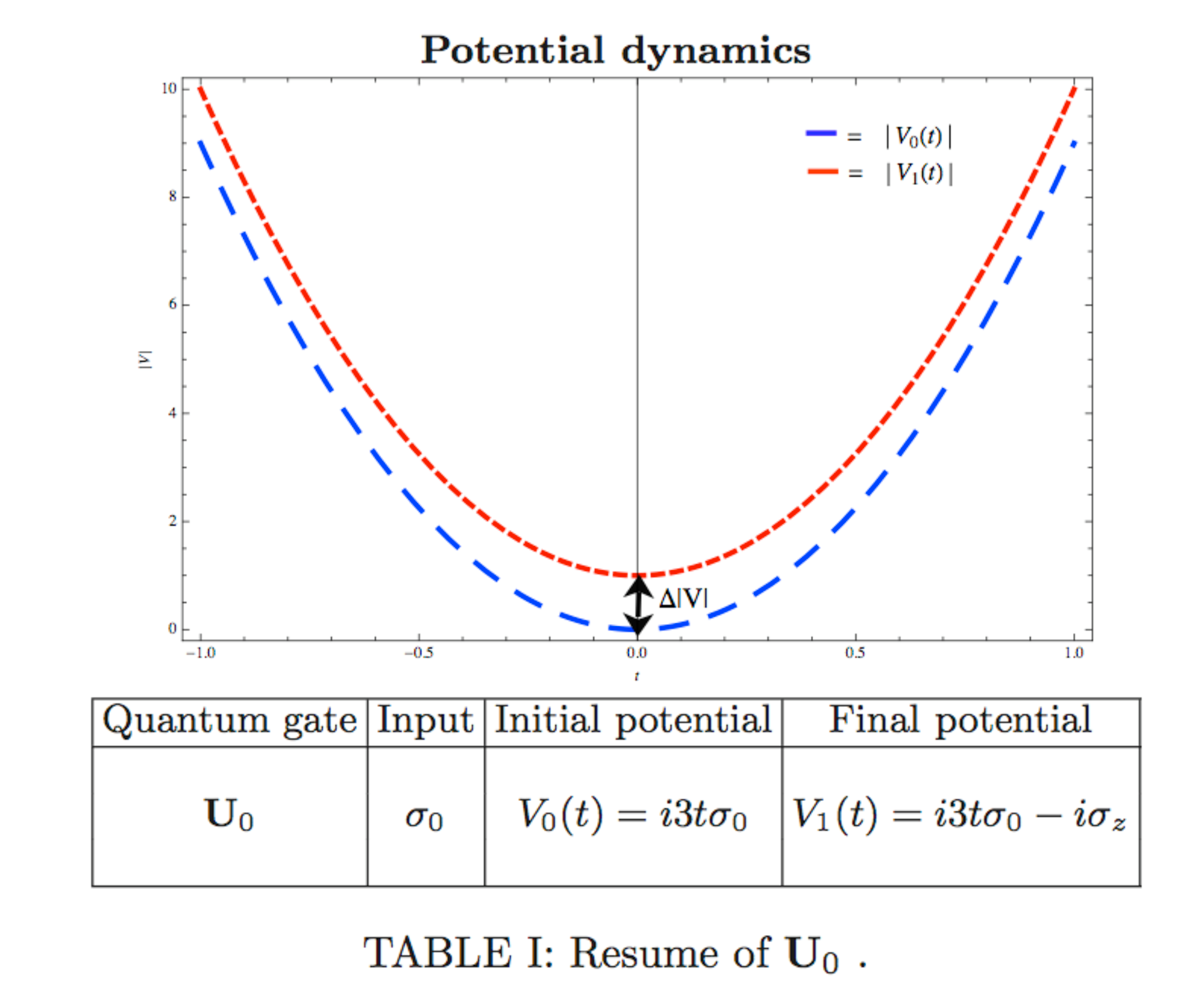}
\caption[f1]{\label{1w}}
\end{figure}

The perturbation on a moving relativistic charge qubit due to external electromagnetic fields fits nicely with the use of BBGS-Darboux transformation on one-dimensional Dirac equation in which the potential is represented in a complex four-vector form. The mathematical method in this case takes benefit from the \textit{vector-quantum gates duality} of Pauli matrices: Pauli matrices can be used as the basis vector or the elementary quantum gate. To reconcile this discrepancy, we argue that the type of quantum gate is related to the direction of Dirac perturbation potential in which is in the Lorentz force form. Mathematically, the perturbation occurs if there is an action of Darboux transformation on the system as mentioned in the Theorem 2 in the next Section. Under this scheme, it is necessary to obtain a quantum logic gate by tuning the perturbation potential, i.e., in our work depends on the direction of the external electromagnetic fields on a relativistic charge qubit moving along $z$-axis. One of our results, for instance, is shown in Fig. (\ref{1w}), i.e., the resume to generate \textbf{IDENTITY}-gate. This phenomenon is similar to the case of quantum state transition due to carrier-photon scattering event in \textit{intraband} of semiconductor.  

\section{Theorems and Proofs}

\textit{${\textbf {Theorem 1}}$. Let $\{V,\Psi\}$ represents a physical system and ${\mathcal D (\sigma_{i})}[N]$ 
is a BBGS-Darboux transformation operator. The open-loop control mechanism can be constructed, where the BBGS-Darboux transformation operator can be assumed as a controller, $\{V,\Psi\}$ as the initial system, $\{V[N],\Psi[N]\}$ as the final system, and the eigenvalues $\varepsilon_{N}(\sigma_{i})$ are the output variables.} \cite{trisetyarso01}

$\textit{Proof}$.  Let  $\{V,\Psi\}$ is the initial condition of a physical system, in which the initial eigenvalues $\varepsilon_{0}$ are belong to the system. 

The action of BBGS-Darboux transformation on this system is ${\mathcal D (\sigma_{i})}[N]$ 
 $\{V,\Psi\}$ =  $\{V[N],\Psi[N]\}$, in which the eigenvalues $\varepsilon_{N}(\sigma_{i})$ are belong to the new eigenstates $\Psi[N]$.

The complete proofs of the above theorem are given in Ref. \cite{trisetyarso01}. In this work, we only consider $N=1$ or one fold Darboux transformations.

Let us consider another scheme: initially the particle is at rest and has the Dirac potential $V_{0}$ and in quantum state $|i\rangle$. This set of Dirac potential and state, $\{V_{0},|i\rangle\}$, belongs to the following one-dimensional stationary Dirac equations
\begin{equation}
\label{eq1}
\hat{h}_{0}\Psi =\varepsilon_{0}\Psi.
\end{equation}
where $\hat{h}_{0}$ = ($i\sigma_{z}\frac{d}{dt}$+$V_{0}(t)$), $V_{0}(t)=\sum_{j}\sigma_{j}(f_{0}(t))_{j}$, and $\varepsilon_{0}$ is a constant. 

$\textbf{Theorem~2}$. \textit{Suppose the initial potential is $V_{0}(t)=\sum_{i}\sigma_{i}(f_{0}(t))_{i}$. The one fold BBGS-Darboux transformations on the equation (\ref{eq1}) at which the final potential is $V_{1}(t)=V_{0}(t)+\Delta V, where~\Delta V= -i\sigma_{z}\textbf{U}_{i}$,  is a suffice condition for} $\hat{\mathcal L}(\{\textbf{U}_{i}\})=\{\textbf{U}_{i}\}.$ $(V_{0}(t))_{0}$ \textit{and} $(V_{1}(t))_{0}$ \textit{are the vector variables.}\cite{trisetyarso02}

\textit{Proof}. Consider 
\begin{equation}
\label{eq2}
\hat{\mathcal L}(\textbf{U}_{i})\hat{h}_{0}\Psi=\hat{h}_{1}\hat{\mathcal L}(\textbf{U}_{i})\Psi.
\end{equation}

One can find

\begin{subequations}
\label{eq7s}
\begin{align}
\label{eq7sa}
\sigma_{0}(V_{0}(t)-V_{0}(t))~~~~~~~~~~~~~~~~~\nonumber
\\+i[\textbf{U}_{i},\sigma_{z}](V_{0}(t)-\beta_{i}(t)-1)=0~
\end{align}
\begin{align}
\label{eq7sb}
i\sigma_{z}\textbf{U}_{i}(\dot{\beta}_{i}(t)-\dot{V}_{0}(t)-\alpha_{i})~~~~~~~~~~~~~~~~~~~~~~~~~~~~~~~~~~~~~~~~~~~~~~~~~~~\nonumber
\\+\alpha_{i}\sigma_{0}(V_{0}(t)-V_{0}(t))~~~~~~~~~~~~~~~~~~~~~~~~~~~~~~~~~~~~~~~~~~~~~~~~~~~~~~~~\nonumber
\\+ (V_{0}(t)-\beta_{0}(t))(V_{0}(t)-(V_{0}(t))\textbf{U}_{i}~~~~~~~~~~~~~~~~~~~~~~~~~~~~~~~~~~~~~~~~~~~~~~~\nonumber
\\+i\sigma_{z}((\beta_{0}(t)-V_{0}(t)-\dot{\alpha}_{i}(t))+\sigma_{0}\dot{V}_{0}(t).~~~~~~~~~~~~~~~~~~~~~~~~~~~~~~~~~~~~~~~~~~
\end{align}
\end{subequations}
\noindent

\textit{Acknowledgements}- This work was supported in part by Grant-in-Aid for ScientiÞc Research
by MEXT, Specially Promoted Research No. 18001002, in part by Special Coordination
Funds for Promoting Science and Technology, and in part by Directorate of Research and Society Service (DP2M) of Telkom Institute of Technology. We also would like to thank Prof. Kohei M.
Itoh, Rodney Van Meter Ph.D, Munawar Riyadi (UTM, Malaysia), and also Ismail Rusli (Telkom Polytechnic) for fruitful discussion.

\begin{IEEEbiography}[{\includegraphics[width=1in,height=1.25in,clip,keepaspectratio]{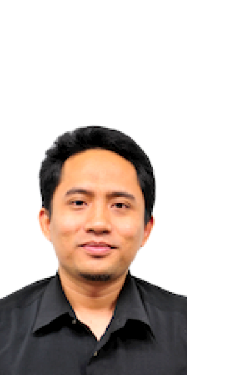}}]
{Agung Trisetyarso, Ph.D} was born in DKI Jakarta, on the 19th of May 1976. He is a Lecturer in Graduate School of Informatics Engineering from Institut Teknologi Telkom (IT Telkom), Bandung, Indonesia. He received the B.S. degree in physics from the Institut Teknologi Bandung (ITB), Indonesia, in 2000, then the M.S. also in physics from ITB in June 2002. By March 2011, he received his Ph.D. degrees in applied physics and physico-informatics from the \textit{Keio Gijuku Daigaku} or Keio University in Tokyo, Japan. He was a \textit{Monbukagakusho} and also \textit{Keio Leading-edge Laboratory} recipient (2007-2010) during the Ph.D study. Currently, he is interested in doing research in quantum computing, machine learning, intelligent system, theoretical computer science and physics, social media, smart city, and also internet of things. 
\end{IEEEbiography}

\end{document}